\documentstyle[sprocl]{article}

\newcommand{\be}{\begin{equation}}
\newcommand{\ee}{\end{equation}}
\newcommand{\ba}{\begin{array}}
\newcommand{\ea}{\end{array}}
\newcommand{\bena}{\begin{eqnarray}}
\newcommand{\eena}{\end{eqnarray}}
\newcommand{\bdis}{\begin{displaymath}}
\newcommand{\edis}{\end{displaymath}}
\newcommand{\bit}{\begin{itemize}}
\newcommand{\eit}{\end{itemize}}
\newcommand{\ben}{\begin{enumerate}}
\newcommand{\een}{\end{enumerate}}

\newcommand{\re}{{\cal I \! \!R}}

\newcommand{\zpcn}{{\cal Z }_c ( \beta,N )}
\newcommand{\isn}{\Omega_N (v)}

\newcommand{\vb}{v \! \! \! \!  -}

\newcommand{\Si}{\Sigma}

\newcommand{\ngV}{\| \nabla V \|}

\newcommand{\de}{\partial}

\newcommand{\lpt}{\left(}
\newcommand{\rpt}{\right)}

\newcommand{\lps}{\langle}
\newcommand{\rps}{\rangle}

\newcommand{\unv}{\frac{1}{\| \nabla V \|}}

\input{psfig}

\def\Journal#1#2#3#4{{#1} {\bf #2}, #3 (#4)}

\def\PRL{\em Phys. Rev. Lett.}
\def\PRE{{\em Phys. Rev.} E}

\def\be{\begin{equation}}
\def\ee{\end{equation}}
\def\bea{\begin{eqnarray}}
\def\eea{\end{eqnarray}}

\begin{document}

\title{TOPOLOGY AND PHASE TRANSITIONS: TOWARDS A PROPER
       MATHEMATICAL DEFINITION OF FINITE N TRANSITIONS}

\author{MARCO PETTINI,~~ROBERTO FRANZOSI\footnote{Dipartimento di Fisica, 
Politecnico di Torino, Italy},~~LIONEL SPINELLI\footnote{CPT-CNRS, Luminy,
Marseille, France}}

\address{Osservatorio Astrofisico di Arcetri, Largo E. Fermi 5, 
   \\ 50125 Firenze, Italy \\E-mail: pettini@arcetri.astro.it} 




\maketitle\abstract{A new point of view about the deep origin of 
thermodynamic phase transitions
is sketched. The main idea is to link the appearance of phase transitions to
some major topology change of suitable submanifolds of phase space instead of
linking them to non-analyticity, as is usual in the Yang-Lee and in the
Dobrushin-Ruelle-Sinai theories. In the new framework a new possibility 
appears to 
properly define a mathematical counterpart of phase transitions also at finite
number of degrees of freedom. This is of prospective interest  to the study of
those systems that challenge the conventional approaches,
as is the case of phase transitions in nuclear clusters.}

\section{Introduction}
The current mathematical definitions of thermodynamical phase transitions are
 based on
the loss of analyticity of thermodynamical observables. This is to some extent
suggested, though not implied, by the experimental relations among macroscopic
observables. This conflicts with the analytic character of the statistical 
weights  that are attributed by any ensemble in statistical mechanics to the 
microscopic configurations. Thus, as it has been proved by the Onsager 
solution of the 2D Ising model and by the Yang-Lee theorem, the only way
to eliminate this conflict is to consider the limit $N\rightarrow\infty$ 
(thermodynamic limit) \cite{LeeYang}.

Obviously phase transitions in Nature occur at finite $N$, but it is commonly 
argued that -- for macroscopic objects -- $N$ is so large that it can be 
considered ``infinite'' from a physical point of view.
However, by looking at a small and embroidery-like snowflake that melts into 
a drop of water one can wonder why such a phenomenon should ever be explained
only in terms of the infinite $N$ limit. 

Moreover, there is a growing experimental evidence that phase transitions can 
occur also at {\it finite and small N}, i.e. with $N\ll$ Avogadro number. 
Some examples of small $N$ systems undergoing phase transitions are: 
{\it i)} nuclear clusters as well as atomic and molecular clusters;
{\it ii)} nano and mesoscopic systems; {\it iii)} polymers and proteins;
{\it iv)} small drops of quantum fluids (BEC, superfluids and superconductors).

Whence the prospective interest of a new mathematical characterization of
thermodynamical phase transitions which, instead of resorting to the loss of
analyticity of macroscopic observables, might naturally encompass also finite
$N$ systems. The search for a broader mathematical definition of phase 
transitions is also of potential interest to the treatment of other important
topics in statistical physics, as is the case of amorphous and disordered
systems (like glasses and spin-glasses), or for a better understanding of
phase transitions in the microcanonical ensemble, for first-order phase 
transitions, and so on.
\subsection{Heuristic arguments}\label{heurist}
Everything here refers to classical Hamiltonian systems with continuous 
variables and described by a standard Hamiltonian  
\be
H [p=(p_1,\dots,p_N),q=(q_1,\dots,q_N)] = \sum_i \frac{1}{2} p_i^2 + V(q)~~.
\label{ham}
\ee
As a consequence of a systematic study of the dynamical counterpart of 
thermodynamical phase transitions, performed by numerically solving the 
Hamilton equations of motion of (\ref{ham}), it has been found that Lyapunov 
exponents display ``singular'' energy and temperature patterns at the 
transition point \cite{cccp}. Thus, since in a 
differential-geometric description of Hamiltonian chaos Lyapunov exponents are
seen as probes of the geometry of certain submanifolds of configuration space,
it has been conjectured that a phase transition could be due to some major
geometrical, and possibly topological, change in the support of the 
statistical measures \cite{cccp,top,physrep}. 
This is to say that it is on the basis of the just mentioned work that we 
have been led to formulate the following argument. 
 
Let us consider the {\it canonical} configurational partition function 
\be
Z_c(\beta ,N) = \int_{\re^N} d^N\! \!q \ e^{-\beta V(q)}=
\int_0^{+\infty} dv \ e^{-\beta v} 
{\int_{\Sigma_v^{N-1}} \frac{d\sigma}{ \Vert \nabla V \Vert} }
\label{zeta}
\ee
where a co-area formula has been used to unfold the structure integrals
\be
\isn \equiv \int_{\Sigma_v^{N-1}}\frac{d\sigma}{\Vert \nabla V \Vert} 
\label{structure}
\ee
i.e an infinite collection of purely geometric integrals on $\Si_v^{N-1}$, 
the {\it equipotential hypersurfaces} of configuration space defined by
$\Si_v^{N-1} \equiv \{ q \in \re^N | V(q) = v \} \subset \re^N$.

If we consider the {\it microcanonical} ensemble, the basic mathematical 
object is the phase space volume
\[
\Omega (E)=\int_0^Ed \eta \  \Omega^{(-)}(E-\eta ) \ \int d^Np \ 
\delta (\sum_i \frac{1}{2} p_i^2 - \eta )
\]
where
\be
\Omega^{(-)}(E-\eta )=\int d^Nq\ \Theta [V(q) - (E-\eta )] = 
\int_0^{E-\eta}dv\ \int_{\Sigma_v}\frac{d\sigma}{\Vert \nabla V \Vert}
\ee
whence
\be
\Omega (E)=\int_0^Ed \eta \ \frac{(2\pi\eta)^{N/2}}{\eta\Gamma(\frac{N}{2})}
\int_0^{E-\eta}dv\  \int_{\Sigma_v}\frac{d\sigma}{\Vert \nabla V \Vert}~,
\label{omega}   
\ee
also here, as in the above decomposition of $Z_c(\beta ,N)$, the only 
non-trivial objects are the structure integrals (\ref{structure}).

Once the microscopic interaction potential $V(q)$ is given, the
configuration space of the system is automatically foliated into the family
$\{ \Sigma_v\}_{v\in{\re}}$ of equipotential hypersurfaces independently of
any statistical measure we may wish to use.
Now, from standard statistical mechanical arguments we know that the larger 
is the number $N$ of particles the closer to some $\Sigma_v$ are the
microstates that significantly contribute to the statistical averages 
of thermodynamic observables. At large $N$, and at any given value of the
inverse temperature $\beta$, the effective support of the canonical measure
is  narrowed very close to a single $\Sigma_v=\Sigma_{v(\beta_c)}$; similarly, 
in the microcanonical ensemble, the fluctuations of potential and kinetic
energies tend to vanish at increasing $N$ so that the effective contributions
to $\Omega(E)$ come from a close neighborhood of a $\Sigma_v=\Sigma_{v(E_c)}$.

Now, 
the ``{\it topological conjecture}'' consists in assuming that some suitable
change of the topology  of the $\{ \Sigma_v\}$,
occurring at some $v_c=v_c(\beta_c)$ (or $v_c=v_c(E_c)$),  is the deep
origin of the singular behavior of thermodynamic observables  at a
phase transition; (by change of topology we mean that  $\{ \Sigma_v\}_{v<v_c}$
are {\it not diffeomorphic} to the $\{ \Sigma_v\}_{v>v_c}$).
In other words, the claim is that the canonical and microcanonical measures
must ``feel'' a big and sudden change -- if any -- of the topology
of the equipotential hypersurfaces of their underlying supports, the 
consequence
being the appearance of the typical signals of a phase transition, i.e.
almost singular energy or temperature dependences of the averages of
appropriate observables.
The larger is $N$ the narrower is the effective support of the measure and
hence the sharper can be the mentioned signals. Eventually, in the 
$N\rightarrow\infty$ limit this sharpening will lead to non-analyticity.

\section{A theorem about topology and phase transitions}
We have recently proved a theorem stating that {\it topological changes of
the hypersurfaces $\Sigma_v$ are necessarily at the origin of phase 
transitions} \cite{bob-tesi,lionel-tesi}. It applies to physical systems 
described by short-range potentials $V$, bounded below, of the general form:
\[
        V(\{q\}) = \sum_{\lps \alpha, \gamma \rps} 
        V_0(\| \vec{q}_\alpha - \vec{q}_\gamma  \|) +
        \sum_{\alpha} \Phi (\| \vec{q}_\alpha \|)~~.
\]
The theorem is proved in the reversed formulation: {\sl if} the surfaces 
{$\Si_{\vb}$} with ${\vb }=V/N \in I= [{\vb}_m,{\vb}_M]$ 
are {\it diffeomorphic}, {\sl then} {\it no phase transition} will occur in 
the corresponding temperature interval  $[\beta ({\vb}_m),\beta ({\vb}_M)]$.
The proof is lenghty and rather complicated but it proceeds along a logically
simple path. Diffeomorphicity of the $\Sigma_v$, after the ``non-critical 
neck theorem'' in Morse theory, implies the absence of critical points of $V$,
i.e. $\nabla V\neq 0$.
{In the absence of Morse critical points:}
{
        \[
  \frac{d^k}{ dv^k} \left( \int_{\Sigma_v} \frac{d\sigma}{\ngV} \right) 
        = \int_{\Sigma_{v}} D^k \left(\unv \right) d\sigma 
        \]}
{where}
{
\[
  D^1({\Vert\nabla V\Vert^{-1}})= 2{\Vert\nabla V\Vert^{-2}}M_1
- {\Vert\nabla V\Vert^{-3}}{\triangle V }
\]}
and where $M_1$ is the trace of the shape operator of the hypersurface 
$\Sigma_v$.
{There is an {operator algebra} to generate
the powers {$D^k$} so that, being ${S}_N({\vb}) =  \frac{1}{N} \ln \Omega_N 
\lpt v({\vb}) \rpt$ the microcanonical configurational entropy, it is 
possible to show that }{
\[
 \sup_{N,\vb\in I} S_N( {\vb}) < \infty~~;~~~
  \sup_{N,\vb\in I} \frac{\de^k 
        S_N}{\de {\vb}^k}({\vb}) < \infty~, k=1,\dots ,4
\]
that is: if the $\Sigma_v$ are diffeomorphic then $S_N({\vb})$ is 
{\it uniformely convergent} in ${\cal C}^3(I)$ as $N\rightarrow\infty$, and, 
by using the Legendre transform relationship 
$S_N(\vb) = f_N(\beta) + \beta \cdot \vb +\ {o}(N)$  
between microcanonical configurational entropy and the canonical free energy
$f_N(\beta) = \frac{1}{N} \ln \zpcn$, this implies that -- in the 
$N\rightarrow\infty$ limit --  
the canonical configurational free energy is
$f_\infty (\beta)\in{\cal C}^2(I)$, i.e. at least twice differentiable,  
thus there are neither first nor second 
order phase transitions according to their standard definition.
There is not a one-to-one correspondence between phase transitions and
topology changes of the $\Si_v$, the latters are necessary but not sufficient.
Sufficiency conditions, to point out the special class of topology changes 
that give rise to phase transitions, are based on relations 
like \cite{bob-tesi,lionel-tesi}
\[
\frac{d\isn}{dv} \simeq 
\int_{\Si_v} \frac{M_1}{\ngV} 
        \frac{d \sigma}{\ngV} \sim 
c(v)[Vol(S_1^{(N-1)})]^{1/N}\left[\sum_{i=0}^N b_i(\Si_v)\right]^\frac{1}{N}
\]
that bridge thermodynamics and topology; here $b_i(\Si_v)$ are the
Betti numbers (cohomological invariants) of $\Si_v$. It turns out that a 
``{\it second order}'' topology change, 
i.e. a sudden change in the way of changing 
of topology as a function of $v$, is sufficient to entail a first or a second
order phase transition. Such {\it topological discontinuities can exist and
can be detected at any finite} $N$, though they yield non-analyticity of 
thermodynamic observables only when the support of the statistical measure 
indefinitely shrinks with $N\rightarrow\infty$. In other words, we have here 
the possibility of properly defining phase transitions also at finite 
$N$, whose detection, direct or indirect, can be performed 
through quantities that 
probe the topology of the $\Si_v$. It is an interesting and open question
how to make the link with other approaches tackling the finite $N$ transitions
in a macroscopic phase space of thermodynamic variables \cite{Gross}.
\section{A direct numerical confirmation}
\label{sec_euler}
The family of $\{\Sigma_v\}_{v\in\re}$ associated with a
$\varphi^4$ model on a
$d$-dimensional lattices $Z^d$ with $d=1,2$ has been used \cite{euler} 
for a numerical 
check of the scenario sketched in the previous section. The model is described
by 
\begin{equation}
V=\sum_{i\in Z^d}\left( - \frac{\mu^2}{2}
q_i^2 + \frac{\lambda}{4\!} q_i^4 \right) +
\sum_{\langle ik\rangle\in
Z^d}\frac{1}{2}J (q_i-q_k)^2
\label{potfi4}
\end{equation}
$\langle ik\rangle$ stands for nearest-neighbor sites, and in $d=2$
it undergoes a second order phase transition. 
In order to directly probe if and how the topology change -- in the sense of
a breaking of {\it diffeomorphicity} of the surfaces $\Sigma_v$ -- is actually
the counterpart of a phase transition,  a {\it diffeomorphism invariant} 
has to be computed. This is a very challenging task because of the high 
dimensionality of the manifolds involved. One possibility is afforded by
the Gauss-Bonnet-Hopf theorem that relates 
the Euler characteristic $\chi (\Sigma_v)$
with the total Gauss-Kronecker curvature $K_G$ of the manifold 
\begin{equation}
\chi (\Sigma_u)  = \gamma \int_{\Sigma_v} K_G \,d\sigma
\label{gaussbonnet}
\end{equation}
which is valid in general for even dimensional hypersurfaces of euclidean 
spaces ${\re}^N$, and where $\gamma =2/Vol({\bf S}^n_1)$ is twice the 
inverse of the volume of an $n$-dimensional sphere of unit radius, and 
$d\sigma$  is the invariant volume measure induced from ${\re}^N$.
\begin{figure} 
\centerline{\psfig{file=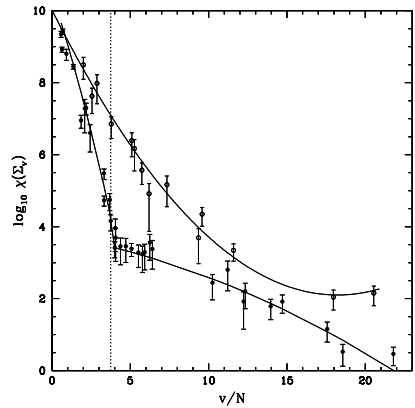,height=6cm,clip=true}}  
\caption{Euler characteristic $\chi (\Sigma_v)$ for 1-$d$ and 2-$d$ 
$\varphi^4$ lattice models.
Open circles:  1-$d$ case, $N = 49$; full circles: 2-$d$ case, $N =
7 \times 7$. The
vertical dotted line, computed separately for larger $N$, accurately
locates the phase transition. Data are from Ref.[8].} 
\label{fig_chi(v)}  
\end{figure}
In Fig.\ \ref{fig_chi(v)} $\chi (\Sigma_v)$ is 
reported {\it vs} $v$: the $1d$ case gives a ``smooth'' pattern
of $\chi (v)$, whereas the $2d$ case yields a cusp-like shaped $\chi (v)$
at the phase transition point. There is here a direct evidence of a major and 
very sharp ``{\it second order}'' topological transition that underlies the 
phase transition, it is
also remarkable that with a very small lattice of $N=7\times 7$ sites such a
sharp signal would never be obtained through standard thermodynamic 
observables \cite{jpa98}.

\end{document}